\documentclass[prl,a4paper,twocolumn,superscriptaddress,showpacs]{revtex4}

\usepackage{amsmath}
\usepackage{amssymb}
\usepackage{textcomp}
\usepackage{upgreek}
\usepackage{units}
\usepackage[pdftex]{graphicx}
\usepackage{multirow}
\usepackage{microtype}
\usepackage{enumitem}

\usepackage{txfonts}

\renewcommand{\vec}[1]{\ensuremath{\boldsymbol{#1}}}

\begin{document}

\title{Tunable magnon Weyl points in ferromagnetic pyrochlores}

\author{Alexander Mook}
\affiliation{Max-Planck-Institut f\"ur Mikrostrukturphysik, D-06120 Halle (Saale), Germany}
\author{J\"urgen Henk}
\affiliation{Institut f\"ur Physik, Martin-Luther-Universit\"at Halle-Wittenberg, D-06099 Halle (Saale), Germany}
\author{Ingrid Mertig}
\affiliation{Max-Planck-Institut f\"ur Mikrostrukturphysik, D-06120 Halle (Saale), Germany}
\affiliation{Institut f\"ur Physik, Martin-Luther-Universit\"at Halle-Wittenberg, D-06099 Halle (Saale), Germany}
\begin{abstract}
The dispersion relations of magnons in ferromagnetic pyrochlores with Dzyaloshinskii-Moriya interaction is shown to possess Weyl points, i.\,e., pairs of topological nontrivial crossings of two magnon branches with opposite topological charge. As a consequence of their topological nature, their projections onto a surface are connected by magnon arcs, thereby resembling closely Fermi arcs of electronic Weyl semimetals. On top of this, the positions of the Weyl points in reciprocal space can be tuned widely by an external magnetic field: rotated within the surface plane, the Weyl points and magnon arcs are rotated as well; tilting the magnetic field out-of-plane shifts the Weyl points toward the center $\overline{\Gamma}$ of the surface Brillouin zone.  The theory is valid for the class of ferromagnetic pyrochlores, i.\,e., three-dimensional extensions of topological magnon insulators on kagome lattices. In this Letter, we focus on the $(111)$ surface, identify candidates of established ferromagnetic pyrochlores which apply to the considered spin model, and suggest experiments for the detection of the topological features.  
\end{abstract}


\date{\today}

\maketitle

\paragraph{Introduction.} Ferromagnetic pyrochlores attracted attention with the experimental discovery of the magnon Hall effect  \cite{Onose2010}: a temperature gradient causes a magnon heat flow in transverse direction. This transverse transport is explained by a Berry curvature in reciprocal space \cite{Katsura2010,Matsumoto2011,Matsumoto2011a}, which is introduced by the Dzyaloshinskii-Moriya (DM) interaction \cite{Dzyaloshinsky58,Moriya60}. In addition, the Chern numbers of magnon bulk bands are nonzero, and in accordance with the bulk-boundary correspondence \cite{Hatsugai1993,Hatsugai1993a} topological magnons are found at the edges of two-dimensional kagome lattices \cite{Zhang2013,Mook14b}. Hence, systems featuring topological magnon states are dubbed `topological magnon insulators' (TMIs) \cite{Zhang2013}, because they exhibit many features of electronic topological insulators \cite{Hasan10}. We recall that a pyrochlore lattice can be viewed as an alternating stacking of kagome planes along its $[111]$ direction.

In this Letter, we predict that ferromagnetic pyrochlores exhibit features of another important class of topological nontrivial systems, namely electronic Weyl semimetals \cite{Wan2011,Xu2015}. Their magnon dispersion relations possess Weyl points, i.\,e., touching points of two otherwise gapped magnon branches. A pair of Weyl points is found on a line in reciprocal space which is along an external magnetic field; these Weyl points possess opposite topological charges of $\pm 1$.

At a surface---here, the $(111)$ surface is studied exemplarily---magnon surface states connect the surface-projected Weyl points; because the associated constant-energy cuts are open they are analogs of Fermi arcs in electronic Weyl semimetals and obey the bulk-boundary correspondence just as kagome TMIs do \cite{Mook14b}. On top of this, these arcs turn out to be easily tunable: upon rotating the magnetic field within the surface plane they follow the likewise rotated Weyl points. An out-of-plane rotation reduces the length of the arcs until they collapse at the center $\overline{\Gamma}$ of the surface Brillouin zone (when the field is perpendicular to the surface). This strong dependence on an external parameter calls for experimental verification.

We recall that recently Weyl points have been predicted in breathing pyrochlore lattices without DM interaction but with a noncollinear ground state \cite{Li2016}. However, the present model relies on a ferromagnetic ground state and on the DM interaction; it is thus a natural extension of TMIs on kagome lattices \cite{Zhang2013,Mook14b} to three dimensions. 

\begin{figure}
	\centering
	\includegraphics[width=0.95\columnwidth]{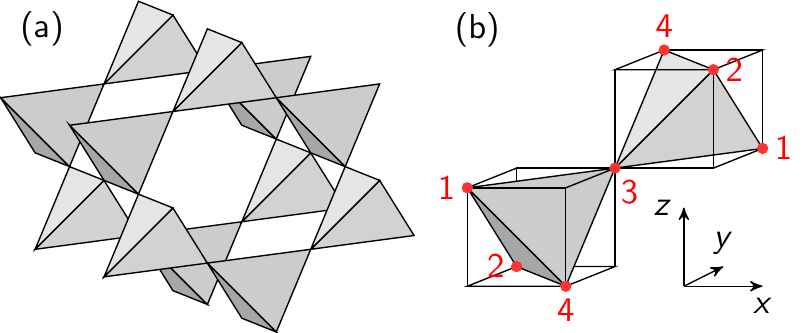}
	\caption{Pyrochlore lattice. 
		(a) Part of the crystal structure. (b) Four basis sites ($1$--$4$) with the nearest-neighbor DM vectors given by 
		$\vec{D}_{12} = D/\sqrt{2} \left( 0 , -1 , -1 \right)$,
		$\vec{D}_{13} = D/\sqrt{2} \left( -1 , 1 , 0 \right)$,
		$\vec{D}_{14} = D/\sqrt{2} \left( 1 , 0 , 1 \right)$,
		$\vec{D}_{23} = D/\sqrt{2} \left( 1 , 0 , -1 \right)$,
		$\vec{D}_{24} = D/\sqrt{2} \left( -1 , -1 , 0 \right)$, and
		$\vec{D}_{43} = D/\sqrt{2} \left( 0 , -1 , 1 \right)$.
	}
	\label{fig:pyro}
\end{figure}

\paragraph{Model and spin-wave analysis.} The pyrochlore lattice is a face-centered cubic (fcc) lattice of corner-sharing tetrahedra, with four atoms in its basis [Fig.~\ref{fig:pyro}(a)]. It lacks inversion symmetry with respect to the midpoints of bonds and, thus, features DM interactions. Moriya's symmetry rules \cite{Moriya60} indicate that the DM vectors $\vec{D}_{ij}$ are perpendicular to the bond that links site $i$ with site $j$; they are situated at the faces of cubes that enclose tetrahedra [Fig.~\ref{fig:pyro}(b)] \cite{Elhajal2005,Kotov2005,Onose2010}.

The minimal magnetic Hamiltonian which includes isotropic symmetric exchange $J_{ij}$, DM interactions $\vec{D}_{ij}$, and a Zeeman term with external magnetic field $\vec{B}$ reads
\begin{align}
	H & = - \sum_{ij} J_{ij} \vec{s}_i \cdot \vec{s}_j
	     + \sum_{ij} \vec{D}_{ij} \cdot \left( \vec{s}_i \times \vec{s}_j \right)
	     - \sum_{i} \vec{B} \cdot \vec{s}_i ,
	    \label{eq:Hamiltonian}
\end{align}
where $\vec{s}_i$ is the spin operator at site $i$. For $J_{ij} > 0$ the ferromagnetic state is the ground state with order parameter $\vec{n} = \vec{B}/B$; the collinearity is stable against the DM interaction \cite{Onose2010}. In the following, the (tiny) rigid energy shift due to the Zeeman energy is neglected and only $\vec{n}$ is kept for simplicity, that is, the limit $B \rightarrow 0^{+}$ is considered.

As argued in Ref.~\cite{Onose2010}, only the component of $\vec{D}_{ij}$ parallel to $\vec{n}$ contributes to the spin-wave Hamiltonian. The spin-wave approximation is then performed in an orthonormal basis $\{ \vec{l}, \vec{m}, \vec{n} \}$, where ladder operators $s_i^\pm = s_i^l \pm \mathrm{i} s_i^m$ are introduced. An adequate Holstein-Primakoff \cite{Holstein1940} transformation is applied; it maps eq.~\eqref{eq:Hamiltonian} onto a boson model in which vacuum accounts for the ferromagnetic ground state, i.\,e., the absence of magnons. A Fourier transformation of the boson operators then yields the spin-wave Hamiltonian which can be solved without a Bogoliubov transformation due to the ferromagnetic ground state. Allowing for nearest-neighbor interactions with strength $J_\mathrm{N}$ as well as next-nearest-neighbor interactions with strength $J_\mathrm{NN}$, the Hamiltonian has the form \cite{Onose2010}
\begin{align}
	H_{\vec{k}} = s \begin{pmatrix}
		H_0			&	H_{12}		& 	H_{13}		& 	H_{14} \\
		H_{12}^\ast	&	H_0			& 	H_{23}		& 	H_{24} \\
		H_{13}^\ast	&	H_{23}^\ast	& 	H_0			& 	H_{34} \\
		H_{14}^\ast	&	H_{24}^\ast	& 	H_{34}^\ast	& 	H_0
	\end{pmatrix},
	\label{eq:SWHamiltonian}
\end{align}
with $H_0 = 6 \left( J_\mathrm{N} + J_\mathrm{NN} \right)$, and $ H_{ij} = - 2 (J_\mathrm{N} + \mathrm{i} D_{ij}^{\vec{n}}) \cos \left( \vec{k} \cdot \vec{\delta}_{ij}^\mathrm{N} \right) - 2 J_\mathrm{NN} \cos \left( \vec{k} \cdot \vec{\delta}_{ij}^\mathrm{NN}\right) $; here, $D_{ij}^{\vec{n}} \equiv \vec{D}_{ij} \cdot \vec{n}$ and $\vec{\delta}_{ij}^\mathrm{N}$ ($\vec{\delta}_{ij}^\mathrm{NN}$) connects nearest (next-nearest) basis sites $i$ and $j$. To mimic pyrochlore systems like, for example, Lu$_2$V$_2$O$_7$, the spin length $s$ is $1/2$ for all basis sites.

\begin{figure}
	\centering
	\includegraphics[width=0.95\columnwidth]{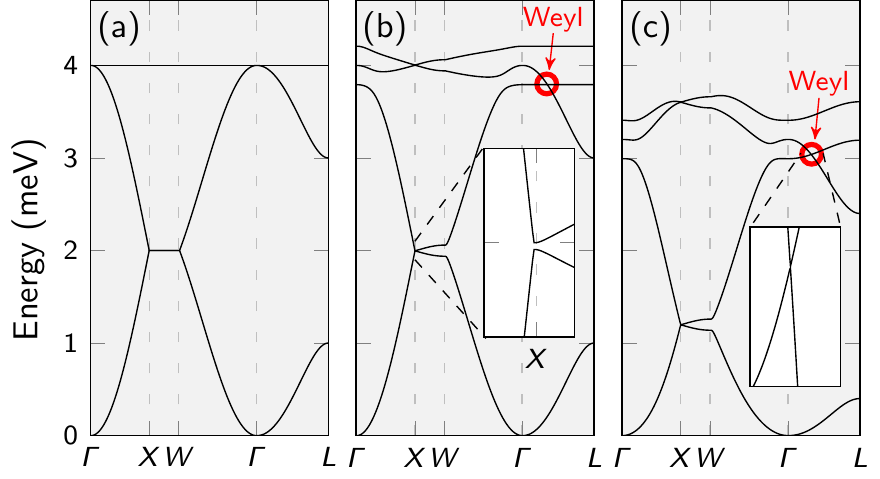}
	\caption{Magnon spectrum of ferromagnetic pyrochlores for (a) $J_\mathrm{N} = \unit[1]{meV}$, $J_\mathrm{NN} = D = 0$, (b) $J_\mathrm{N} = \unit[1]{meV}$,  $J_\mathrm{NN} = 0$, $D = \unit[0.28]{meV}$, and (c) $J_\mathrm{N} = \unit[1]{meV}$, $J_\mathrm{NN} = \unit[-0.1]{meV}$, $D = \unit[0.28]{meV}$. The magnetic field points along $\vec{n} = (1,1,1)/\sqrt{3}$, i\,e.\ along $\Gamma \rightarrow L$. Red circles in (b) and (c) mark Weyl points.}
	\label{fig:bs}
\end{figure}

We now analyze magnon spectra $\varepsilon_{\nu \vec{k}}$ (band index $\nu=1,2,3,4$), calculated from eq.~\eqref{eq:SWHamiltonian}, and the associated Berry curvatures \cite{Berry1984,Zak1989}
\begin{align}
	\vec{\Omega}_{ \nu \vec{k}} &\equiv \mathrm{i}
	\sum_{\mu \neq \nu} 
	\frac{
	\left\langle \vec{u}_{\nu \vec{k}} \left| \partial_{\vec{k}} H_{\vec{k}} \right| \vec{u}_{\mu \vec{k}} \right\rangle \times
	\left\langle \vec{u}_{\mu \vec{k}} \left| \partial_{\vec{k}} H_{\vec{k}} \right| \vec{u}_{\nu \vec{k}} \right\rangle
	}{
	\left( \varepsilon_{\nu \vec{k}} - \varepsilon_{\mu \vec{k}}\right)^2
	},
		\label{eq:curvature-def} 
\end{align}
where $\left| \vec{u}_{\nu \vec{k}} \right\rangle$ is an eigenvector of $H_{\vec{k}}$ with energy $\varepsilon_{\nu\vec{k}}$. Typical results are summarized in panels (a) -- (c) of Fig.~\ref{fig:bs}.
\begin{enumerate}[label=(\alph*)]
\item For $J_\mathrm{NN} = 0$ and $D = 0$, the four magnon bands are not gapped, the third and fourth band are dispersionless and degenerate. The Berry curvature vanishes in the entire Brillouin zone (BZ), the direction $\vec{n}$ of the magnetic field does not affect the spectrum.

\item For $J_\mathrm{NN} = 0$ but $D > 0$, the upper two magnon branches become dispersive and the Berry curavture is nonzero for all bands. The latter is explained by the complex hopping elements $J_\mathrm{N} + \mathrm{i} D_{ij}^{\vec{n}}$ that enter $H_{\vec{k}}$.

The band structure depends sensitively on $\vec{n}$. As long as $\vec{n}$ is \emph{not} within a $\{1\, 0 \,0 \}$ plane, a tiny fundamental gap between the first and second band shows up (inset). 

More strikingly, on any line in reciprocal space through the origin $\Gamma$ and parallel to $\vec{n}$ [$\vec{n} \parallel (\Gamma - L)$ in (b)], the second and the third band cross each other at two $\vec{k}$; the latter lie symmetric to $\Gamma$, their spacing is determined by $\vec{n}$ and $|D|$. As we will show in the following, these band crossings are \emph{Weyl points} (red circle).

\item The Weyl points are robust against $J_\mathrm{NN} \neq 0$ (inset). In particular, for antiferromagnetic next-nearest neighbor coupling ($J_\mathrm{NN} < 0$, small enough to retain the ferromagentic ground state), the Weyl points are located in energy so that no other bulk band has the same energy (for any $\vec{n}$). We choose these parameters because they facilitate the following analysis.
\end{enumerate}

\paragraph{Topology of bulk bands.} In the remainder of this Letter, we present results obtained for the exchange parameters of Fig.~\ref{fig:bs}(c).

An integer Chern number
\begin{align}
	C_{\nu} = \frac{1}{2 \pi} \int_{S} \vec{\Omega}_{\nu \vec{k}} \cdot \tilde{\vec{n}} \, \mathrm{d}S,
\end{align}
is calculated for each band $\nu$; $S$ is a closed and oriented surface in the bulk BZ with surface normal $\tilde{\vec{n}}$.
Letting $\tilde{\vec{n}} = \mathrm{const.}$, $S$ is a 2D slice of the 3D BZ. 
By moving the slice in the BZ, $C_{\nu}(\lambda)$ can be calculated as a function of the position $\lambda$ of the slice as in Ref.~\onlinecite{Hermanns2015}. 

\begin{figure}
	\centering
	\includegraphics[width=0.8\columnwidth]{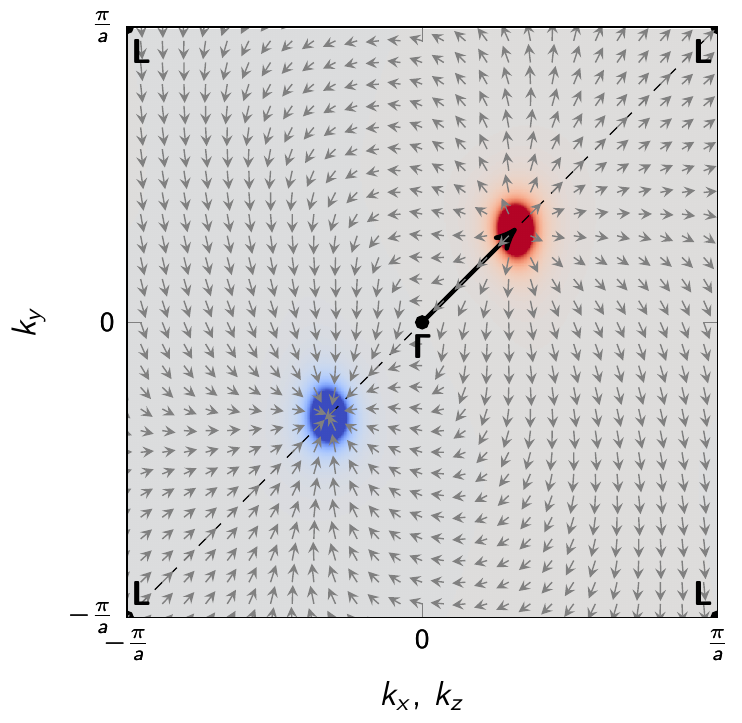}
	\caption{Berry curvature and Weyl points. The normalized dipole vector field $\vec{\Omega}_{2\vec{k}}$ of band $2$ is shown in the $k_{x} = k_{z}$ plane. High-symmetry points $\Gamma$ and $L$ are indicated. The color scale depicts the divergence of the vector field (blue: negative, gray: zero, red: positive); the two Weyl points appear in the center of the blue and red spot, respectively. Parameters as in Fig.~\ref{fig:bs}(c), in particular $\vec{n} = (1,1,1)/\sqrt{3}$, i.\,e.\,, along $\Gamma \rightarrow L$ (indicated by the black arrow).}
	\label{fig:berry-field}
\end{figure}

It is $C_{1}(\lambda) = -\operatorname{sgn}(D)$ a constant function (here, $\tilde{\vec{n}}=\vec{n}$), because there is a fundamental band gap that separates the lowest from the other bands [cf.\ the inset in Fig.~\ref{fig:bs}(b)]. 
In contrast, $C_2(\lambda)$ and $C_3(\lambda)$ are not globally constant, but only piece-wise constant, because the bands touch each other at the Weyl points. The latter are monopoles of the Berry curvature vector field, which is evident from the denominator in eq.~\eqref{eq:curvature-def}. To prove that the band crossings are indeed Weyl points we show the Berry curvature vector field of band $2$ (Fig.~\ref{fig:berry-field}). There are two monopoles that appear as source (red spot, offset from $\Gamma$ in the direction of the magnetic field) and sink (blue spot) of the vector field, providing clear evidence that the Weyl points have opposite topological charge $q^\mathrm{top}_{2}$, which is identical to $C_{2}$, where $S$ is a sphere enclosing one of the Weyl points. Numerical integration yields $q_{2}^{\mathrm{top}} = + 1$ for the `red' (source) and $q_{2}^{\mathrm{top}} = -1$ for the `blue' (sink) Weyl point in Fig.~\ref{fig:berry-field}.

Summarizing at this point, we have identified Weyl points in the bulk band structure of pyrochlore systems whose positions in reciprocal space can be tuned by an external magnetic field. A prominent feature of electronic Weyl semimetals are Fermi arcs which are surface states that connect projections of Weyl points onto the surface BZ. In the following we show that pyrochlore systems exhibit the magnon analogs of the (electronic) Fermi arcs, that is magnon arcs.

\paragraph{Surface states.} The surface magnon dispersion is analyzed in terms of the spectral density $N_{p}(\varepsilon, \vec{k})$ which is calculated by Green function renormalization \cite{Henk1993}. We evaluate a semi-infinite geometry, thereby avoiding finite-size effects that appear in slab calculations.

The main idea of the renormalization is as follows. For a chosen surface, the pyrochlore lattice is decomposed into principal layers (PLs) which are parallel to that surface. The principal layers have to be chosen so that the Hamiltonian matrix of the semi-infinite system comprises only interactions within a PL and among adjacent PLs. In the infinite set of equations for the PL-resolved Green function the inter-PL interactions can iteratively be reduced (renormalized), so that the entire Hamilton matrix becomes block diagonal, allowing to compute the Green function blocks $G_{pp}$. The latter yield the spectral densities
\begin{align}
	N_{p}(\varepsilon, \vec{k}) & = - \frac{1}{\pi} \lim_{\eta \rightarrow 0^+} 
	\mathrm{Im} \left[ \operatorname{Tr} G_{pp} (\varepsilon + \mathrm{i} \eta, \vec{k}) \right]
\end{align}
of PL $p$. Hence, we have access to both the bulk spectral density ($p = \infty$) and to that of any other PL, in particular that of the surface ($p = 0$). A finite $\eta$ introduces broadening; here: $\eta = \unit[0.001]{meV}$.

In the following, we study the $(111)$ surface as an example and choose quite large a DM interaction $D$ to provide a clear picture. We would like to stress that the discussion is qualitatively valid for all surfaces and all $D$.

\begin{figure*}
	\centering
	\includegraphics[width=0.95\textwidth]{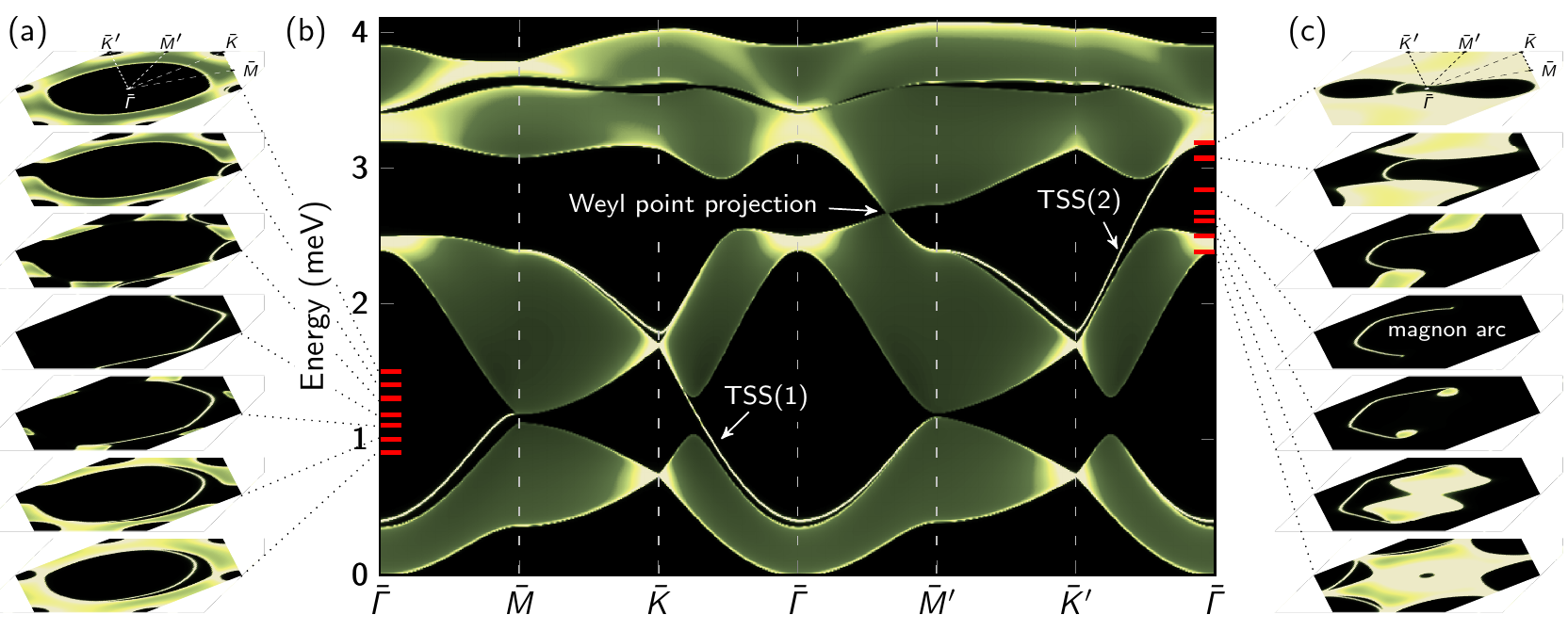}
	\caption{Magnons at the $(111)$ surface of a ferromagnetic pyrochlore. The surface spectral density $N_{0}(\varepsilon, \vec{k})$ is shown as color scale (black: zero; white: maximum). Bulk magnons appear as broad features, surface states as sharp lines. (a) and (c) show constant-energy cuts through the entire surface Brillouin zone for energies indicated by red lines in (b). (b) Spectral density along high-symmetry directions of the surface Brillouin zone. The projection of the Weyl points and two topological surface states (TSS) are indicated.	Parameters as for Fig.~\ref{fig:bs}(c), except $D=\unit[0.5]{meV}$.}
	\label{fig:top-weyl-way}
\end{figure*}

The $(111)$ surface of the pyrochlore lattice is a kagome lattice, the resulting surface BZ is a hexagon [Fig.~\ref{fig:top-weyl-way}(a)]. Note that a magnetic field with an in-plane component breaks the rotational symmetry of the surface BZ\@. The magnetic field is completely in-plane along $[11\overline{2}]$; hence, the projections of the Weyl points are situated on the line $\overline{M}'-\overline{\Gamma}-\overline{M}'$ [Fig.~\ref{fig:top-weyl-way}(b)].

The bulk bands appear as broad features in (b). The gaps between band $1$ and $2$ as well band $2$ and $3$ are bridged by two topological surface states, TSS(1) and TSS(2). The latter obey the bulk-boundary correspondence \cite{Hatsugai1993,Hatsugai1993a,Mook14b,Mook15b}.

We recall that the Chern number of band $1$ is $-1$; accordingly, the winding number of the gap between band $1$ and $2$ equals $-1$ as well (the winding number is the sum of all Chern numbers of the bands below the considered gap). Since the winding number dictates the number of topological nontrivial surface states, there has to be one topological surface state: TSS(1). For kagome systems confer Refs.~\onlinecite{Zhang2013} and~\onlinecite{Mook14b}.

The bulk-boundary correspondence also holds for TSS(2), since the Weyl points have topological charges (Chern numbers), but we now show that TSS(1) and TSS(2) differ qualitatively. Therefore, we present constant-energy cuts of the surface spectrum that  cover the band gaps in which TSS(1) and TSS(2) are situated [(a) and (c)]. Both surface states are easily identified as bright lines clinging to the extended bulk features.

At $\varepsilon = \unit[1.18]{meV}$ bulk states do not contribute to the surface spectral density and only TSS(1) is visible [center figure in (a)]. Apparently, TSS(1) forms a closed line when considering the periodicity of the surface BZ\@.

Considering TSS(2), a similar scenario takes place at $\varepsilon = \unit[2.67]{meV}$ which is the energy of the Weyl points [center figure of (c)]. Instead of a closed line, we find a \emph{magnon arc} that connects the projections of the two Weyl points of opposite topological charge. Thus, pyrochlore ferromagnets host the magnon pendant to the Fermi arcs in electronic Weyl semimetals. We recall that magnon arcs were not found in (two-dimensional) kagome lattices.

We now show that the Weyl points and the associated magnon arcs can be shifted---i.\,e., tuned---by the magnetic field $\vec{B}$. The energy of the Weyl points is not affected by the rotation of $\vec{B}$; hence, all of the constant-energy contours discussed in what follows are at the same energy ($\unit[2.67]{meV}$).

\begin{figure*}
	\centering
	\includegraphics[width=0.95\textwidth]{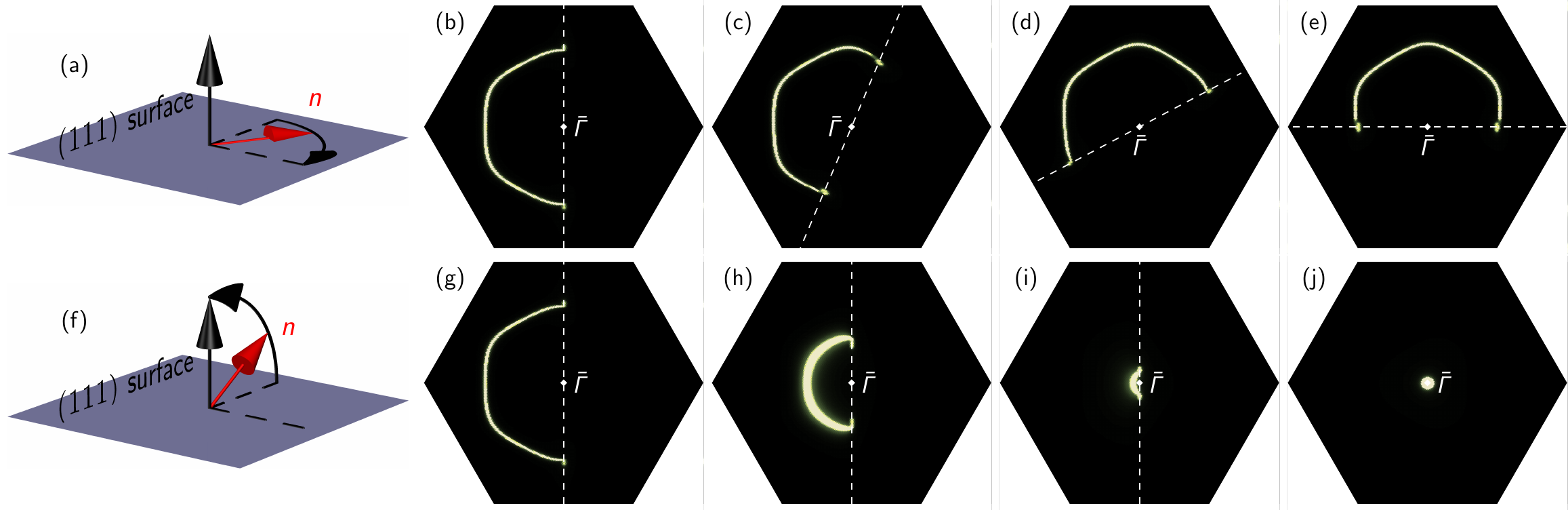}
	\caption{Tuning magnon arcs in a pyrochlore ferromagnet by an external magnetic field. Top row (a)--(e): effect of rotating the magnetic field within the (111) surface, as sketched in (a), on the magnon arc (b)--(e). Bottom row (f)--(j): as the top panels but for rotating the magnetic field out-of-plane, as sketched in (f). In (j) the field is perpendicular to the surface and the magnon arc `collapses'. The direction of the magnetic field projected onto the $(111)$ surface is indicated by white dashed lines. Parameters as for Fig.~\ref{fig:top-weyl-way}, except $\vec{n}$ which is varied.}
	\label{fig:top-weyl}
\end{figure*}

By rotating the field within the surface plane, depicted in Fig.~\ref{fig:top-weyl}(a), the Weyl points follow the magnetic field and can be rotated arbitrarily. Consequently, the magnon arc trails the Weyl point projections and is rotated likewise [(b)--(e)]. The arc is not rotated rigidly: it shows the largest distances from $\overline{\Gamma}$ along $\overline{\Gamma}-\overline{M}$ as well as $\overline{\Gamma}-\overline{M}'$ directions, irrespective of the magnetic  field's azimuth. A sign change of either $D$ or $\vec{n}$ would change the signs of the Berry curvature and of the topological charges; the magnon arc would be reflected about the direction of the magnetic field.

Rotating the magnetic field from in-plane to out-of-plane (bottom row in Fig.~\ref{fig:top-weyl}), the Weyl point projections are shifted toward $\overline{\Gamma}$, thereby reducing the length of the respective magnon arc. Eventually, the arc `collapses' when the magnetic field points along the surface normal [Fig.~\ref{fig:top-weyl}(j)].

\paragraph{Experimental considerations.} The pyrochlore oxides Lu$_2$V$_2$O$_7$, In$_2$Mn$_2$O$_7$, and Ho$_2$V$_2$O$_7$, all of which exhibit the magnon Hall effect \cite{Ideue12}, are the most promising candidates for experimental detection of magnon Weyl points. The first two are modeled very well by the Hamiltonian in eq.~\eqref{eq:Hamiltonian}. For Lu$_2$V$_2$O$_7$, the ratio $D  / J_{\mathrm{N}}$ of DM interaction to exchange interaction has been determined recently, with values of $0.32$ \cite{Onose2010}, $0.18$ \cite{Mena2014}, $0.07$ \cite{Riedl2016}, and $0.05$ \cite{Xiang11}. Since the DM interaction determines the distance between the Weyl points and $\Gamma$ in reciprocal space, a search for Weyl points can help to identify the exact ratio. The  tunability of the Weyl points and magnon arcs can be exploited, for example, in inelastic neutron scattering experiments \cite{Mena2014}: the shifts of the bulk band crossings upon variation of the external magnetic field can be traced. 
By probing a fixed line in reciprocal space through the origin, say $\vec{q}$, one will see a band gap closing and reopening at one point on $\vec{q}$ upon evolution of the field's azimuth. The closing, i.\,e.\,, the occurrence of Weyl points, coincides with the alignment of the field and $\vec{q}$.

Electron energy loss spectroscopy, which is sensitive to the surface \cite{Zakeri2013}, could be applied for the detection of the topological magnon surface states, in particular, the magnon arcs. 
Keeping in mind that the surface spectral density of the bulk states does not change under inversion of the magnetic field, but that of the topological surface states is reflected, a difference experiment could be conducted: subtracting the results of oppositely magnetized samples yields clear evidence of surface states.

Furthermore, since the transverse thermal conductivities of Lu$_2$V$_2$O$_7$ and In$_2$Mn$_2$O$_7$ differ in sign \cite{Ideue12}, it is likely that their DM constants $D$ have opposite signs as well. Therefore, the magnon arcs of the two systems should roughly be mirror images for the same experimental setup (given that the exchange constants and the absolute value of the DM constants of the two crystals do not differ drastically).

\paragraph{Conclusion.} Ferromagnetic pyrochlores feature magnon Weyl points that can easily be tuned by an external magnetic field. Thus, the class of topologically nontrivial systems which comprises topological magnon insulators is extended to, loosely speaking, `magnon Weyl semimetals'. The latter consists of noncollinear breathing pyrochlores, as reported earlier \cite{Li2016}, and ferromagnetic pyrochlores, as predicted in this Letter. The effect of magnon Weyl points on magnon transport of both spin and heat as well the formation of topological interface modes \cite{Mook15a,Mook15b} appear worth investigated in the future.

\paragraph{Acknowledgments.} This work is supported by SPP 1666 of Deutsche Forschungsgemeinschaft (DFG).

\bibliography{short,newrefs}

\begin{thebibliography}{28}
\expandafter\ifx\csname natexlab\endcsname\relax\def\natexlab#1{#1}\fi
\expandafter\ifx\csname bibnamefont\endcsname\relax
  \def\bibnamefont#1{#1}\fi
\expandafter\ifx\csname bibfnamefont\endcsname\relax
  \def\bibfnamefont#1{#1}\fi
\expandafter\ifx\csname citenamefont\endcsname\relax
  \def\citenamefont#1{#1}\fi
\expandafter\ifx\csname url\endcsname\relax
  \def\url#1{\texttt{#1}}\fi
\expandafter\ifx\csname urlprefix\endcsname\relax\def\urlprefix{URL }\fi
\providecommand{\bibinfo}[2]{#2}
\providecommand{\eprint}[2][]{\url{#2}}

\bibitem[{\citenamefont{Onose et~al.}(2010)\citenamefont{Onose, Ideue, Katsura,
  Shiomi, Nagaosa, and Tokura}}]{Onose2010}
\bibinfo{author}{\bibfnamefont{Y.}~\bibnamefont{Onose}},
  \bibinfo{author}{\bibfnamefont{T.}~\bibnamefont{Ideue}},
  \bibinfo{author}{\bibfnamefont{H.}~\bibnamefont{Katsura}},
  \bibinfo{author}{\bibfnamefont{Y.}~\bibnamefont{Shiomi}},
  \bibinfo{author}{\bibfnamefont{N.}~\bibnamefont{Nagaosa}}, \bibnamefont{and}
  \bibinfo{author}{\bibfnamefont{Y.}~\bibnamefont{Tokura}},
  \bibinfo{journal}{Science} \textbf{\bibinfo{volume}{329}},
  \bibinfo{pages}{297} (\bibinfo{year}{2010}),
  \urlprefix\url{http://dx.doi.org/10.1126/science.1188260}.

\bibitem[{\citenamefont{Katsura et~al.}(2010)\citenamefont{Katsura, Nagaosa,
  and Lee}}]{Katsura2010}
\bibinfo{author}{\bibfnamefont{H.}~\bibnamefont{Katsura}},
  \bibinfo{author}{\bibfnamefont{N.}~\bibnamefont{Nagaosa}}, \bibnamefont{and}
  \bibinfo{author}{\bibfnamefont{P.~A.} \bibnamefont{Lee}},
  \bibinfo{journal}{Phys.\ Rev.\ Lett.} \textbf{\bibinfo{volume}{104}},
  \bibinfo{pages}{066403} (\bibinfo{year}{2010}),
  \urlprefix\url{http://dx.doi.org/10.1103/PhysRevLett.104.066403}.

\bibitem[{\citenamefont{Matsumoto and
  Murakami}(2011{\natexlab{a}})}]{Matsumoto2011}
\bibinfo{author}{\bibfnamefont{R.}~\bibnamefont{Matsumoto}} \bibnamefont{and}
  \bibinfo{author}{\bibfnamefont{S.}~\bibnamefont{Murakami}},
  \bibinfo{journal}{Phys.\ Rev.\ B} \textbf{\bibinfo{volume}{84}},
  \bibinfo{pages}{184406} (\bibinfo{year}{2011}{\natexlab{a}}),
  \urlprefix\url{http://dx.doi.org/10.1103/PhysRevB.84.184406}.

\bibitem[{\citenamefont{Matsumoto and
  Murakami}(2011{\natexlab{b}})}]{Matsumoto2011a}
\bibinfo{author}{\bibfnamefont{R.}~\bibnamefont{Matsumoto}} \bibnamefont{and}
  \bibinfo{author}{\bibfnamefont{S.}~\bibnamefont{Murakami}},
  \bibinfo{journal}{Phys.\ Rev.\ Lett.} \textbf{\bibinfo{volume}{106}},
  \bibinfo{pages}{197202} (\bibinfo{year}{2011}{\natexlab{b}}),
  \urlprefix\url{http://dx.doi.org/10.1103/PhysRevLett.106.197202}.

\bibitem[{\citenamefont{Dzyaloshinsky}(1958)}]{Dzyaloshinsky58}
\bibinfo{author}{\bibfnamefont{I.}~\bibnamefont{Dzyaloshinsky}},
  \bibinfo{journal}{J. Phys.\ Chem.\ Sol.} \textbf{\bibinfo{volume}{4}},
  \bibinfo{pages}{241} (\bibinfo{year}{1958}), ISSN \bibinfo{issn}{0022-3697}.

\bibitem[{\citenamefont{Moriya}(1960)}]{Moriya60}
\bibinfo{author}{\bibfnamefont{T.}~\bibnamefont{Moriya}},
  \bibinfo{journal}{Phys.\ Rev.} \textbf{\bibinfo{volume}{120}},
  \bibinfo{pages}{1} (\bibinfo{year}{1960}).

\bibitem[{\citenamefont{Hatsugai}(1993{\natexlab{a}})}]{Hatsugai1993}
\bibinfo{author}{\bibfnamefont{Y.}~\bibnamefont{Hatsugai}},
  \bibinfo{journal}{Physical Review B} \textbf{\bibinfo{volume}{48}},
  \bibinfo{pages}{11851} (\bibinfo{year}{1993}{\natexlab{a}}),
  \urlprefix\url{http://dx.doi.org/10.1103/PhysRevB.48.11851}.

\bibitem[{\citenamefont{Hatsugai}(1993{\natexlab{b}})}]{Hatsugai1993a}
\bibinfo{author}{\bibfnamefont{Y.}~\bibnamefont{Hatsugai}},
  \bibinfo{journal}{Physical Review Letters} \textbf{\bibinfo{volume}{71}},
  \bibinfo{pages}{3697–3700} (\bibinfo{year}{1993}{\natexlab{b}}), ISSN
  \bibinfo{issn}{0031-9007},
  \urlprefix\url{http://dx.doi.org/10.1103/PhysRevLett.71.3697}.

\bibitem[{\citenamefont{Zhang et~al.}(2013)\citenamefont{Zhang, Ren, Wang, and
  Li}}]{Zhang2013}
\bibinfo{author}{\bibfnamefont{L.}~\bibnamefont{Zhang}},
  \bibinfo{author}{\bibfnamefont{J.}~\bibnamefont{Ren}},
  \bibinfo{author}{\bibfnamefont{J.-S.} \bibnamefont{Wang}}, \bibnamefont{and}
  \bibinfo{author}{\bibfnamefont{B.}~\bibnamefont{Li}},
  \bibinfo{journal}{Phys.\ Rev.\ B} \textbf{\bibinfo{volume}{87}},
  \bibinfo{pages}{144101} (\bibinfo{year}{2013}),
  \urlprefix\url{http://dx.doi.org/10.1103/PhysRevB.87.144101}.

\bibitem[{\citenamefont{Mook et~al.}(2014)\citenamefont{Mook, Henk, and
  Mertig}}]{Mook14b}
\bibinfo{author}{\bibfnamefont{A.}~\bibnamefont{Mook}},
  \bibinfo{author}{\bibfnamefont{J.}~\bibnamefont{Henk}}, \bibnamefont{and}
  \bibinfo{author}{\bibfnamefont{I.}~\bibnamefont{Mertig}},
  \bibinfo{journal}{Phys. Rev. B} \textbf{\bibinfo{volume}{90}},
  \bibinfo{pages}{024412} (\bibinfo{year}{2014}),
  \urlprefix\url{http://link.aps.org/doi/10.1103/PhysRevB.90.024412}.

\bibitem[{\citenamefont{Hasan and Kane}(2010)}]{Hasan10}
\bibinfo{author}{\bibfnamefont{H.}~\bibnamefont{Hasan}} \bibnamefont{and}
  \bibinfo{author}{\bibfnamefont{C.}~\bibnamefont{Kane}},
  \bibinfo{journal}{Rev.\ Mod.\ Phys.} \textbf{\bibinfo{volume}{82}},
  \bibinfo{pages}{3045} (\bibinfo{year}{2010}).

\bibitem[{\citenamefont{Wan et~al.}(2011)\citenamefont{Wan, Turner, Vishwanath,
  and Savrasov}}]{Wan2011}
\bibinfo{author}{\bibfnamefont{X.}~\bibnamefont{Wan}},
  \bibinfo{author}{\bibfnamefont{A.~M.} \bibnamefont{Turner}},
  \bibinfo{author}{\bibfnamefont{A.}~\bibnamefont{Vishwanath}},
  \bibnamefont{and} \bibinfo{author}{\bibfnamefont{S.~Y.}
  \bibnamefont{Savrasov}}, \bibinfo{journal}{Phys. Rev. B}
  \textbf{\bibinfo{volume}{83}}, \bibinfo{pages}{205101}
  (\bibinfo{year}{2011}),
  \urlprefix\url{http://link.aps.org/doi/10.1103/PhysRevB.83.205101}.

\bibitem[{\citenamefont{Xu et~al.}(2015)\citenamefont{Xu, Alidoust, Belopolski,
  Yuan, Bian, Chang, Zheng, Strocov, Sanchez, Chang et~al.}}]{Xu2015}
\bibinfo{author}{\bibfnamefont{S.-Y.} \bibnamefont{Xu}},
  \bibinfo{author}{\bibfnamefont{N.}~\bibnamefont{Alidoust}},
  \bibinfo{author}{\bibfnamefont{I.}~\bibnamefont{Belopolski}},
  \bibinfo{author}{\bibfnamefont{Z.}~\bibnamefont{Yuan}},
  \bibinfo{author}{\bibfnamefont{G.}~\bibnamefont{Bian}},
  \bibinfo{author}{\bibfnamefont{T.-R.} \bibnamefont{Chang}},
  \bibinfo{author}{\bibfnamefont{H.}~\bibnamefont{Zheng}},
  \bibinfo{author}{\bibfnamefont{V.~N.} \bibnamefont{Strocov}},
  \bibinfo{author}{\bibfnamefont{D.~S.} \bibnamefont{Sanchez}},
  \bibinfo{author}{\bibfnamefont{G.}~\bibnamefont{Chang}},
  \bibnamefont{et~al.}, \bibinfo{journal}{Nature Physics}
  \textbf{\bibinfo{volume}{11}}, \bibinfo{pages}{748–754}
  (\bibinfo{year}{2015}), ISSN \bibinfo{issn}{1745-2481},
  \urlprefix\url{http://dx.doi.org/10.1038/nphys3437}.

\bibitem[{\citenamefont{Li et~al.}(2016)\citenamefont{Li, Li, Kim, Balents, Yu,
  and Chen}}]{Li2016}
\bibinfo{author}{\bibfnamefont{F.-Y.} \bibnamefont{Li}},
  \bibinfo{author}{\bibfnamefont{Y.-D.} \bibnamefont{Li}},
  \bibinfo{author}{\bibfnamefont{Y.-B.} \bibnamefont{Kim}},
  \bibinfo{author}{\bibfnamefont{L.}~\bibnamefont{Balents}},
  \bibinfo{author}{\bibfnamefont{Y.}~\bibnamefont{Yu}}, \bibnamefont{and}
  \bibinfo{author}{\bibfnamefont{G.}~\bibnamefont{Chen}}
  (\bibinfo{year}{2016}), \eprint{1602.04288}.

\bibitem[{\citenamefont{Elhajal et~al.}(2005)\citenamefont{Elhajal, Canals,
  Sunyer, and Lacroix}}]{Elhajal2005}
\bibinfo{author}{\bibfnamefont{M.}~\bibnamefont{Elhajal}},
  \bibinfo{author}{\bibfnamefont{B.}~\bibnamefont{Canals}},
  \bibinfo{author}{\bibfnamefont{R.}~\bibnamefont{Sunyer}}, \bibnamefont{and}
  \bibinfo{author}{\bibfnamefont{C.}~\bibnamefont{Lacroix}},
  \bibinfo{journal}{Phys.\ Rev.\ B} \textbf{\bibinfo{volume}{71}},
  \bibinfo{pages}{094420} (\bibinfo{year}{2005}),
  \urlprefix\url{http://dx.doi.org/10.1103/PhysRevB.71.094420}.

\bibitem[{\citenamefont{Kotov et~al.}(2005)\citenamefont{Kotov, Elhajal,
  Zhitomirsky, and Mila}}]{Kotov2005}
\bibinfo{author}{\bibfnamefont{V.~N.} \bibnamefont{Kotov}},
  \bibinfo{author}{\bibfnamefont{M.}~\bibnamefont{Elhajal}},
  \bibinfo{author}{\bibfnamefont{M.~E.} \bibnamefont{Zhitomirsky}},
  \bibnamefont{and} \bibinfo{author}{\bibfnamefont{F.}~\bibnamefont{Mila}},
  \bibinfo{journal}{Phys. Rev. B} \textbf{\bibinfo{volume}{72}},
  \bibinfo{pages}{014421} (\bibinfo{year}{2005}),
  \urlprefix\url{http://link.aps.org/doi/10.1103/PhysRevB.72.014421}.

\bibitem[{\citenamefont{Holstein and Primakoff}(1940)}]{Holstein1940}
\bibinfo{author}{\bibfnamefont{T.}~\bibnamefont{Holstein}} \bibnamefont{and}
  \bibinfo{author}{\bibfnamefont{H.}~\bibnamefont{Primakoff}},
  \bibinfo{journal}{Phys. Rev.} \textbf{\bibinfo{volume}{58}},
  \bibinfo{pages}{1098} (\bibinfo{year}{1940}),
  \urlprefix\url{http://link.aps.org/doi/10.1103/PhysRev.58.1098}.

\bibitem[{\citenamefont{Berry}(1984)}]{Berry1984}
\bibinfo{author}{\bibfnamefont{M.~V.} \bibnamefont{Berry}},
  \bibinfo{journal}{Proc.\ R.\ Soc.\ London A} \textbf{\bibinfo{volume}{392}},
  \bibinfo{pages}{45} (\bibinfo{year}{1984}),
  \urlprefix\url{http://dx.doi.org/10.1098/rspa.1984.0023}.

\bibitem[{\citenamefont{Zak}(1989)}]{Zak1989}
\bibinfo{author}{\bibfnamefont{J.}~\bibnamefont{Zak}}, \bibinfo{journal}{Phys.
  Rev. Lett.} \textbf{\bibinfo{volume}{62}}, \bibinfo{pages}{2747}
  (\bibinfo{year}{1989}),
  \urlprefix\url{http://link.aps.org/doi/10.1103/PhysRevLett.62.2747}.

\bibitem[{\citenamefont{Hermanns et~al.}(2015)\citenamefont{Hermanns, O'Brien,
  and Trebst}}]{Hermanns2015}
\bibinfo{author}{\bibfnamefont{M.}~\bibnamefont{Hermanns}},
  \bibinfo{author}{\bibfnamefont{K.}~\bibnamefont{O'Brien}}, \bibnamefont{and}
  \bibinfo{author}{\bibfnamefont{S.}~\bibnamefont{Trebst}},
  \bibinfo{journal}{Phys. Rev. Lett.} \textbf{\bibinfo{volume}{114}},
  \bibinfo{pages}{157202} (\bibinfo{year}{2015}),
  \urlprefix\url{http://link.aps.org/doi/10.1103/PhysRevLett.114.157202}.

\bibitem[{\citenamefont{Henk and Schattke}(1993)}]{Henk1993}
\bibinfo{author}{\bibfnamefont{J.}~\bibnamefont{Henk}} \bibnamefont{and}
  \bibinfo{author}{\bibfnamefont{W.}~\bibnamefont{Schattke}},
  \bibinfo{journal}{Comp.\ Phys.\ Commun.} \textbf{\bibinfo{volume}{77}},
  \bibinfo{pages}{69–83} (\bibinfo{year}{1993}), ISSN
  \bibinfo{issn}{0010-4655},
  \urlprefix\url{http://dx.doi.org/10.1016/0010-4655(93)90038-E}.

\bibitem[{\citenamefont{Mook et~al.}(2015{\natexlab{a}})\citenamefont{Mook,
  Henk, and Mertig}}]{Mook15b}
\bibinfo{author}{\bibfnamefont{A.}~\bibnamefont{Mook}},
  \bibinfo{author}{\bibfnamefont{J.}~\bibnamefont{Henk}}, \bibnamefont{and}
  \bibinfo{author}{\bibfnamefont{I.}~\bibnamefont{Mertig}},
  \bibinfo{journal}{Phys. Rev. B} \textbf{\bibinfo{volume}{91}},
  \bibinfo{pages}{224411} (\bibinfo{year}{2015}{\natexlab{a}}),
  \urlprefix\url{http://link.aps.org/doi/10.1103/PhysRevB.91.224411}.

\bibitem[{\citenamefont{Ideue et~al.}(2012)\citenamefont{Ideue, Onose, Katsura,
  Shiomi, Ishiwata, Nagaosa, and Tokura}}]{Ideue12}
\bibinfo{author}{\bibfnamefont{T.}~\bibnamefont{Ideue}},
  \bibinfo{author}{\bibfnamefont{Y.}~\bibnamefont{Onose}},
  \bibinfo{author}{\bibfnamefont{H.}~\bibnamefont{Katsura}},
  \bibinfo{author}{\bibfnamefont{Y.}~\bibnamefont{Shiomi}},
  \bibinfo{author}{\bibfnamefont{S.}~\bibnamefont{Ishiwata}},
  \bibinfo{author}{\bibfnamefont{N.}~\bibnamefont{Nagaosa}}, \bibnamefont{and}
  \bibinfo{author}{\bibfnamefont{Y.}~\bibnamefont{Tokura}},
  \bibinfo{journal}{Phys.\ Rev.\ B} \textbf{\bibinfo{volume}{85}},
  \bibinfo{pages}{134411} (\bibinfo{year}{2012}),
  \urlprefix\url{http://dx.doi.org/10.1103/PhysRevB.85.134411}.

\bibitem[{\citenamefont{Mena et~al.}(2014)\citenamefont{Mena, Perry, Perring,
  Le, Guerrero, Storni, Adroja, R\"uegg, and McMorrow}}]{Mena2014}
\bibinfo{author}{\bibfnamefont{M.}~\bibnamefont{Mena}},
  \bibinfo{author}{\bibfnamefont{R.~S.} \bibnamefont{Perry}},
  \bibinfo{author}{\bibfnamefont{T.~G.} \bibnamefont{Perring}},
  \bibinfo{author}{\bibfnamefont{M.~D.} \bibnamefont{Le}},
  \bibinfo{author}{\bibfnamefont{S.}~\bibnamefont{Guerrero}},
  \bibinfo{author}{\bibfnamefont{M.}~\bibnamefont{Storni}},
  \bibinfo{author}{\bibfnamefont{D.~T.} \bibnamefont{Adroja}},
  \bibinfo{author}{\bibfnamefont{C.}~\bibnamefont{R\"uegg}}, \bibnamefont{and}
  \bibinfo{author}{\bibfnamefont{D.~F.} \bibnamefont{McMorrow}},
  \bibinfo{journal}{Phys. Rev. Lett.} \textbf{\bibinfo{volume}{113}},
  \bibinfo{pages}{047202} (\bibinfo{year}{2014}),
  \urlprefix\url{http://link.aps.org/doi/10.1103/PhysRevLett.113.047202}.

\bibitem[{\citenamefont{Riedl et~al.}(2016)\citenamefont{Riedl, Guterding,
  Jeschke, Gingras, and Valenti}}]{Riedl2016}
\bibinfo{author}{\bibfnamefont{K.}~\bibnamefont{Riedl}},
  \bibinfo{author}{\bibfnamefont{D.}~\bibnamefont{Guterding}},
  \bibinfo{author}{\bibfnamefont{H.~O.} \bibnamefont{Jeschke}},
  \bibinfo{author}{\bibfnamefont{M.~J.~P.} \bibnamefont{Gingras}},
  \bibnamefont{and} \bibinfo{author}{\bibfnamefont{R.}~\bibnamefont{Valenti}}
  (\bibinfo{year}{2016}), \eprint{arXiv:1604.03472}.

\bibitem[{\citenamefont{Xiang et~al.}(2011)\citenamefont{Xiang, Kan, H., Lee,
  Wei, and Gong}}]{Xiang11}
\bibinfo{author}{\bibfnamefont{H.~J.} \bibnamefont{Xiang}},
  \bibinfo{author}{\bibfnamefont{E.~J.} \bibnamefont{Kan}},
  \bibinfo{author}{\bibfnamefont{W.~M.} \bibnamefont{H.}},
  \bibinfo{author}{\bibfnamefont{C.}~\bibnamefont{Lee}},
  \bibinfo{author}{\bibfnamefont{S.-H.} \bibnamefont{Wei}}, \bibnamefont{and}
  \bibinfo{author}{\bibfnamefont{X.~G.} \bibnamefont{Gong}},
  \bibinfo{journal}{Phys.\ Rev.\ B} \textbf{\bibinfo{volume}{83}},
  \bibinfo{pages}{174402} (\bibinfo{year}{2011}), \eprint{1010.4606}.

\bibitem[{\citenamefont{Zakeri et~al.}(2013)\citenamefont{Zakeri, Zhang, and
  Kirschner}}]{Zakeri2013}
\bibinfo{author}{\bibfnamefont{K.}~\bibnamefont{Zakeri}},
  \bibinfo{author}{\bibfnamefont{Y.}~\bibnamefont{Zhang}}, \bibnamefont{and}
  \bibinfo{author}{\bibfnamefont{J.}~\bibnamefont{Kirschner}},
  \bibinfo{journal}{Journal of Electron Spectroscopy and Related Phenomena}
  \textbf{\bibinfo{volume}{189}}, \bibinfo{pages}{157–163}
  (\bibinfo{year}{2013}),
  \urlprefix\url{http://dx.doi.org/10.1016/j.elspec.2012.06.009}.

\bibitem[{\citenamefont{Mook et~al.}(2015{\natexlab{b}})\citenamefont{Mook,
  Henk, and Mertig}}]{Mook15a}
\bibinfo{author}{\bibfnamefont{A.}~\bibnamefont{Mook}},
  \bibinfo{author}{\bibfnamefont{J.}~\bibnamefont{Henk}}, \bibnamefont{and}
  \bibinfo{author}{\bibfnamefont{I.}~\bibnamefont{Mertig}},
  \bibinfo{journal}{Phys. Rev. B} \textbf{\bibinfo{volume}{91}},
  \bibinfo{pages}{174409} (\bibinfo{year}{2015}{\natexlab{b}}),
  \urlprefix\url{http://link.aps.org/doi/10.1103/PhysRevB.91.174409}.

\end{thebibliography}
\bibliographystyle{apsrev}

\end{document}